\documentclass[a4paper]{jpconf}
\usepackage{graphicx}
\usepackage{comment}
\usepackage{amsmath,amssymb,bm}
\begin{document}
\title{Defect-Induced Kondo Effect in graphene: Role of Localized State of $\pi$ Electrons}

\author{Taro  Kanao, Hiroyasu Matsuura, and Masao Ogata}
\address{Department of Physics, University of Tokyo, Bunkyo, Tokyo 113-0033, Japan
}

\ead{kanao@hosi.phys.s.u-tokyo.ac.jp}

\begin{abstract}
We discuss a role of the localized $\pi$ orbital, which exists around the defect, on the defect-induced Kondo effect in graphene by a numerical renormalization group study.  
We find that the localized $\pi$ orbital assists this Kondo effect, and the Kondo temperature is sensitive to the broadening of the localized $\pi$ orbital.
Secondly, we focus on the negative magnetoresistance of this Kondo effect.
In the experimental result, it has been shown that the negative magnetoresistance is ten times larger than the usual Kondo effect.
In order to clarify the mechanism of the "magnetic sensitive" Kondo effect, as a first step, we study an orbital magnetic field dependence of the localized $\pi$ orbital by a tight-binding model with a Peierls phase.
We find that as the magnetic field increases, the spectral width of the localized $\pi$ orbital increases and the local DOS at the Fermi level decreases.
Since the Kondo temperature is strongly dependent of the broadening of the localized $\pi$ orbital, it is expected that this Kondo effect is sensitive to the orbital magnetic field as observed in the experiment. 
\end{abstract}

%%%%%%%%%%%%%%%%%%%%%%%%%%%%%%%%%%%
\section{Introduction}\label{sec_kondo_intro}

After the fabrication of graphene, the Kondo effect in graphene has attracted much theoretical attention, because exotic Kondo effects and associated phenomena due to its Dirac conduction electrons are expected, such as multichannel Kondo effects, gate-tunable Kondo effects, and an impurity quantum phase transition between an unscreened and a screened localized moment state [1-6].%~\cite{Kotov2012,Fritz2012,Sengupta2008,Vojta2010,Zhu2010,Haase2011}.
As candidates of the realization of the Kondo effect in graphene, the graphene with magnetic impurities such as Fe or Co has been investigated by DFT calculations~\cite{Fritz2012}.
The possibility of the Kondo effect in these systems has been argued by examining the lattice positions, the spin state of the impurity atom, and the strength and symmetry of the hybridization between the impurity and the conduction electrons.
Experiments on graphene with such magnetic impurities have been attempted, and spectral signatures of the Kondo screening have been observed in scanning tunneling microscope (STM) experiments~\cite{Brar2011}.
In such systems, however, Kondo effect had not been observed in resistivity measurements.

In 2011, the Kondo effect was reported in the resistivity of graphene with defects~\cite{Chen2011}.
The defects were introduced by the irradiation of He$^{+}$ ions with the energy of 500 eV.
The temperature dependence of resistivity showed logarithmic increase with decreasing temperature and saturating behavior at low temperatures.
The temperature dependence can be fitted well by the universal temperature dependence of the Kondo effect.
In the ion-irradiated graphene, the negative magnetoresistance was also observed~\cite{Chen2011}.
Magnetic field is applied perpendicular to the graphene plane.
The contribution of the weak localization is suppressed by the magnetic field $B\sim1$ T, and a magnetoresistance due to another mechanism occurs at higher magnetic fields.
In the case of the Kondo effect, the Zeeman effect of magnetic field can cause such a negative magnetoresistance.
From these observations, the Kondo effect due to magnetic origin was concluded in this system.
To clarify the origin of the Kondo effect, many theoretical studies have been performed [9-15].%~\cite{Kanao2012, Cazalilla2012,Mitchell2013, Jafari2013, Kharitonov2013,Kanao2013,Shirakawa2014}.
On the other hand, it also has been proposed that the electron-electron interaction in the presence of the disorder is responsible for the logarithmic increase of the resistivity at low temperatures~\cite{Jobst2012, Chen2012}.
Although there is the other possible theory, in this paper, we focus on the defect induced Kondo effect.

Until the discovery of the defect induced Kondo effect, the defects in graphene or graphite have been studied as a source of magnetism~\cite{Yazyev2010}.
In graphite, it has been observed that magnetic moments are induced by irradiation of high-energy (2.25 MeV) protons~\cite{Esquinazi2003}.
In the graphene irradiated with ions, the magnetic moments have also been observed~\cite{Nair2012}.
It has been indicated that the defect individually behaves as a localized magnetic moment.
Such defects have been studied in STM experiments~\cite{Ugeda2010, KondoT2010}.
These experiments show that the introduced defects are point defects (or point vacancies).
The defects in graphene have also been investigated with transmission electron microscopes (TEMs)~\cite{Meyer2008}, and reconstruction and deformation of the defect are observed.
Such a point defect in graphene has been studied with the DFT calculations.
First, the lattice structure was investigated~\cite{El-Barbary2003}.
Two possible structures were considered: one is a threefold symmetric structure that preserves the original symmetry, and the other is a twofold symmetric structure in which one of the C-C distances is shortened.
It was clarified that the latter is more stable owning to the Jahn-Teller effect, and it has an out-of-plane displacement, while the former has a planar structure.
Second, the spin state was investigated by using spin-polarized DFT calculations~\cite{Lehtinen2004,Yazyev2007}.
It was shown that the defect is magnetic and the dangling $sp^2$ bond which appears at the defect is responsible for the magnetic moment.

In graphene, effects of defect scattering on the $\pi$ electrons have been studied theoretically, and it has been suggested that the point defect induces a localized state of $\pi$ electrons around the defect whose energy eigenvalue coincides with the energy of the Dirac points [26-29].%~\cite{Wakabayashi2002, Pereira2006, Kumazaki2007, Toyoda2010}.
Such a localized $\pi$ orbital has been observed experimentally with the STM.
In scanning tunneling spectroscopy (STS) experiments, the localized $\pi$ orbital exhibits a sharp peak in local density of states (LDOS) near the defect.

Although the origin of the Kondo effect in graphene with defects seems to be due to the localized magnetic moments on the $sp^2$ orbital at the defects, the role of localized $\pi$ orbital has not been understood in detail.
Recently, although a model where the localized state of $\pi$ electrons is rather artificially added at the defect~\cite{Cazalilla2012,Shirakawa2014} has been proposed, more consideration on the derivation of the model and the analysis are necessary.
In addition, in this defect induced Kondo effect, the anomalous negative magnetoresistance has been observed: 
In the usual Kondo effect, the energy scale of a characteristic magnetic field of the negative magnetoresistance is the same order as that of the Kondo temperature.
On the other hand, experimentally, the characteristic magnetic field seems to be one tenth smaller than the value expected from the Kondo temperature, which means that this Kondo effect is sensitive to the magnetic field.
However, this mechanism has not been understood yet.

Thus, in this paper, we discuss the role of localized $\pi$ orbital on the defect-induced Kondo effect in graphene on the basis of a numerical renormalization group (NRG) study~\cite{Wilson1975}.  
As a result, we find that the localized $\pi$ orbital assists the Kondo effect, and the Kondo temperature is sensitive to the broadening of the localized $\pi$ orbital.
Next, in order to clarify the mechanism of the magnetic sensitive Kondo effect, as a first step, we study an orbital magnetic field dependence of the localized $\pi$ orbital by a tight-binding model with a Peierls phase.
We show that the LDOS on the localized $\pi$ orbital decreases as the magnetic field increases.

In \S.2, we develop an effective model with the localized $\pi$ orbital and transforms the effective model into a one-dimensional representation of the conduction electron states to apply NRG method.
In \S.3, we discuss the effect of the localized $\pi$ orbital by NRG study.
Finally, in \S.4, we discuss an orbital magnetic field dependence of the localized state of $\pi$ electrons by the tight binding model with the Peierls phase.

%%%%%%%%%%%%%%%%%%%%%%%%%%%%%%%%%%%%%%%%
\section{Localized Moment and its Hybridization with Conduction Electrons}
\subsection{Cluster Model for the Defect}\label{sec_kondo_cluster_model}
Here, considering a cluster for the defect, we introduce a model for the system.
Around the point defect, there are three $sp^2$ orbitals $\varphi_i$ which do not form covalent bond, where $i$ indicates one of the three sites ($i=1$, $2$, or $3$) around the defect. 
There are also $\pi$ orbitals on carbon atoms.

First, we consider an isolated cluster that consists of three $sp^2$ orbitals, as a model for the localized state of $sp^2$ orbitals around the defect.
A tight-binding approximation is applied to this cluster. 
Two of the three transfer integrals between the $sp^2$ orbitals are $-h$, and the other is $-h'$  ($h,h'>0$). 
The threefold symmetric structure corresponds to the case with $h=h'$, and the twofold symmetric one to $h'>h$, where it is assumed that the distance between sites $2$ and $3$ is shorter than the others.

\begin{figure}
	\centering
	\includegraphics[width=14cm]{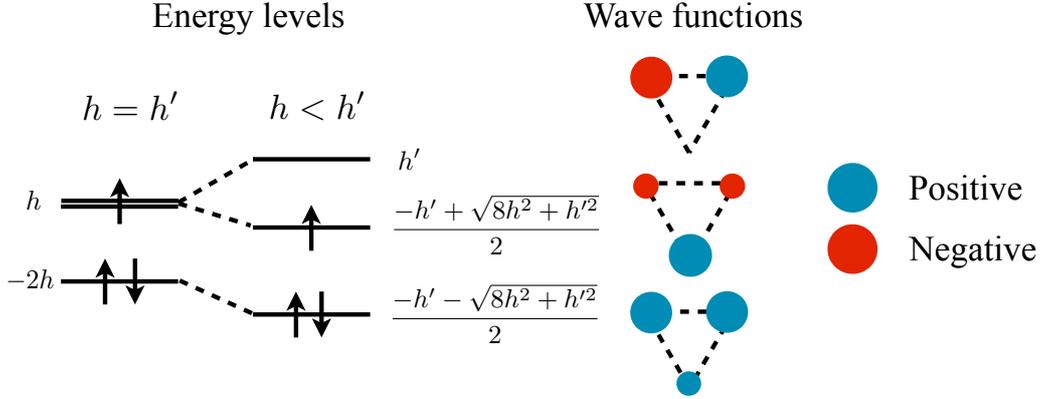}
	\caption{Energy levels and the corresponding wave functions of a cluster model for localized states of $sp^2$ orbitals around a point defect in graphene.}
	\label{fig_kondo_cluster}
\end{figure}

The energy eigenvalues and the corresponding wave functions of this cluster model are
\begin{eqnarray}
	E_3&=&h',\hspace{8em}\psi_3=\frac{1}{\sqrt{2}}(\varphi_2-\varphi_3),\nonumber\\
	E_2&=&\frac{-h'+\sqrt{8h^2+h'^2}}{2},\hspace{1em}\psi_2=C_+\varphi_1-\frac{1}{\sqrt{2}}C_-(\varphi_2+\varphi_3),\nonumber\\
	E_1&=&\frac{-h'-\sqrt{8h^2+h'^2}}{2},\hspace{1em}\psi_1=C_-\varphi_1+\frac{1}{\sqrt{2}}C_+(\varphi_2+\varphi_3),\label{eq_energy_wave_function}
\end{eqnarray}
with
\begin{eqnarray}
	C_\pm&=&\left[\frac{1}{2}\left(1\pm\frac{h'}{\sqrt{8h^2+h'^2}}\right)\right]^{1/2},
\end{eqnarray}
which are depicted in Fig.~\ref{fig_kondo_cluster}.
There are three electrons on these $sp^2$ orbitals. 
Since the number of electrons is odd, there is an unfilled level and nonzero spin appears. 
Interpreting the Kohn-Sham eigenvalues of the DFT calculations~\cite{El-Barbary2003} in terms of these tight-binding energy levels, we semiquantitatively estimate $h'/h$ to be $h'/h\simeq5$.
When $h'/h\simeq5$, $C_\pm$ becomes $C_+\simeq1$, $C_-\ll 1$, and the wave function of the lowest level $\psi_1$ becomes a bonding state $\psi_1=(\varphi_2+\varphi_3)/\sqrt{2}$, while the second state becomes $\psi_2=\varphi_1$.
Thus, $\varphi_1$ becomes the highest occupied molecular orbital of this cluster, and causes a localized magnetic moment.
It should be noted that in a divacancy case, four electrons are present in the $sp^2$ orbitals and the localized spin does not appear. 

\begin{figure}
	\centering
	\includegraphics[width=15cm]{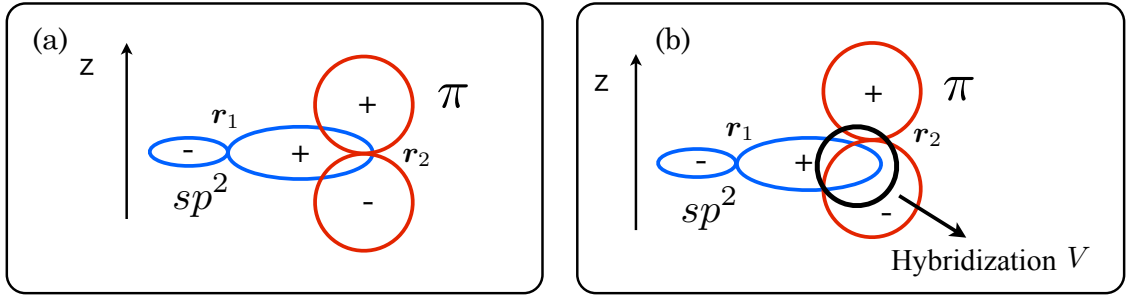}
	\caption{Schematic picture of $sp^2$ (blue) and $\pi$ (red) orbitals on the different carbon atoms at $\bm{r}_1$ and $\bm{r}_2$.
		%The signs ($+$ and $-$) represent those of the orbitals.
		%$z$ axis is perpendicular to the graphene plane.
		(a) Planar structure.
		%The hybridization integral vanishes owning to the sign change of the $\pi$ orbital.
		(b) In the presence of out-of-plane displacement of the carbon atom.
		%Finite hybridization $V$ appears owning to the relative displacement of the $\pi$ orbital.
		}
	\label{fig_kondo_hybridization}
\end{figure}
Second, we consider the hybridization between this localized magnetic moment on the $sp^2$ orbital and the conduction electrons which consist of $\pi$ orbitals.
It depends on the lattice structure of the defect. 
When the defect has the planar structure, there is no hybridization between the $sp^2$ and $\pi$ orbitals, which is schematically shown in Fig.~\ref{fig_kondo_hybridization}(a).
However, with an out-of-plane displacement, finite hybridization between the $sp^2$ orbital $\varphi_i$ and the two $\pi$ orbitals at the other sites $j\neq i$ appears as in Fig.~\ref{fig_kondo_hybridization}(b).
As mentioned above, from the DFT calculations, the twofold symmetric structure with an out-of-plane displacement is suggested to be more stable owing to the Jahn-Teller effect~\cite{El-Barbary2003}.

\begin{figure}
	\centering
		\includegraphics[width=5cm]{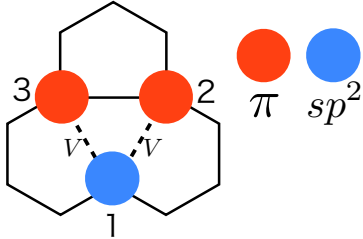}
	\caption{Schematic picture of the electronic state around the defect.
		Two of the three $sp^2$ orbitals ($i=2$, $3$) form a covalent bond.
		The circle at the site $1$ shows the active $sp^2$ orbital and those at the sites 2 and 3 the $\pi$ orbitals.
		$V$ is the amplitude of the hybridization between the active $sp^2$ orbital and the two $\pi$ orbitals.}
	\label{fig_distortion}
\end{figure}

In the following we consider the structure shown in Fig.~\ref{fig_distortion}, where only one $sp^2$ orbital ($i=1$) is active and it hybridizes with the $\pi$ orbitals at sites $2$ and $3$. 
(The bonding state of $sp^2$ orbitals at sites $2$ and $3$ is neglected.)
We assume that the hybridization term is given by 
\begin{eqnarray}
	H_{\rm{hyb}}=\sum_{\sigma=\uparrow\downarrow}\left[V\left(a^\dagger_{2\sigma}+a^\dagger_{3\sigma}\right)d_{\sigma}+\mathrm{h. c. }\right]\label{eq_hybridization},
\end{eqnarray}
where $a_{i\sigma}$ is an annihilation operator of a $\pi$ electron at site $i$ with spin $\sigma=\uparrow,\downarrow$ and $d_\sigma$ is an annihilation operator of an electron on the $sp^2$ orbital at site $i=1$. 
$V$ is the amplitude of hybridization. 
Taking account of the localized nature of the active $sp^2$ orbital in the defect, we assume that the effective Hamiltonian for this orbital becomes 
\begin{eqnarray}
	H_{\rm{def}}=\sum_{\sigma}(\epsilon_{sp^2}-\mu)d^\dagger_{\sigma}d_{\sigma}+Ud^\dagger_{\uparrow}d_{\uparrow}d^\dagger_{\downarrow}d_{\downarrow},\label{eq_defect_hamiltonian}
\end{eqnarray}
where $\epsilon_{sp^2}$, $\mu$, and $U$ are the energy level, chemical potential, and Coulomb interaction on the $sp^2$ orbital, respectively. 
The Coulomb interaction term is essential for describing the Kondo effect. 

The conduction electron states are formed by $\pi$ orbitals in graphene, which are described by the following tight-binding model with a nearest-neighbor transfer integral $t(>0)$, 
\begin{eqnarray}
	H_{\rm{gra}}=-t\sum_{\langle ij\rangle\sigma}{}^{'}\left(a^\dagger_{i\sigma}b_{j\sigma}+\rm{h.c.}\right)-\mu N,\label{eq_graphene_hamiltonian}
\end{eqnarray}
where $b_{j\sigma}$ is an annihilation operator of a $\pi$ electron at site $j$ of the sublattice $B$, and $N$ is the total number of $\pi$ electrons.
In eq.~(\ref{eq_hybridization}), only the $\pi$ orbitals of the sublattice $A$, $a_{i\sigma}$, appear since we have assumed the missing carbon atom to be on the sublattice $B$.
The prime in eq.~(\ref{eq_graphene_hamiltonian}) denotes the presence of the point defect.
The point defect in $\pi$ electrons is represented by the zero transfer integrals between the missing site and the neighboring site.
Here, we ignore the Coulomb interaction in the $\pi$ orbitals.
Although band structure may be modified by the Coulomb interaction in the bulk, the Dirac dispersion is preserved and we assume that eq.~(\ref{eq_graphene_hamiltonian}) represents such renormalized Dirac fermions.
%as mentioned in \S\ref{sec_electron_electron_interaction_in_dirac_fermions}
Finally, the effective Hamiltonian of graphene with the point defect becomes $H=H_{\rm{gra}}+H_{\rm{def}}+H_{\rm{hyb}}$. 

%%%%%%%%%%%%%%%%%%%%%%%%%%%%%%%%%%%%
\subsection{One-Dimensional Representation of the Conduction Electron States}\label{sec_one_dimensional}
To clarify the conduction electron states which are coupled with the $sp^2$ orbital at the defect, and to apply the NRG method, we transform this effective Hamiltonian~(\ref{eq_hybridization}), (\ref{eq_defect_hamiltonian}), and  (\ref{eq_graphene_hamiltonian}) into a one-dimensional representation.
In the usual Kondo problems, a magnetic impurity interacts with conduction electrons at the same site.
In this case, the conduction electron states are expanded by the spherical waves, and states with particular symmetries such as $s$, $p$, or $d$ wave symmetries are coupled with the impurity.
For example, when the interaction between the impurity and the conduction electrons is spherical, only the states with $s$ wave symmetry is coupled with the impurity.
Such states are labeled by the wave vectors in the radial direction.
As a result, a one-dimensional representation is obtained.
However, in the present model, there are two centers: the $sp^2$ orbital at the defect interacts with the $\pi$ orbitals of sites 2 and 3. 
Such a geometry has been treated in the two-impurity Kondo problem~\cite{Jones1987}. 
We follow the method developed for it~\cite{Jones1987,Sakai1992,Affleck1995}, generalizing the method to include the effects of defect scattering on the $\pi$ conduction electrons.

To express the hybridization term (\ref{eq_hybridization}) in a one-dimensional representation, we introduce the following operator for $\pi$ electrons~\cite{Sakai1992,Affleck1995};
\begin{eqnarray}
	a^\dagger_{\varepsilon i\sigma} &=& \sum_k c^\dagger_{k\sigma}\langle k|Ai\rangle\delta\left(\varepsilon-\varepsilon_k\right),\label{eq_operator_energy_site_rep}
\end{eqnarray}
where $\varepsilon_k$ and $|k\rangle$ are the energy eigenvalues and the wave functions of the Hamiltonian~(\ref{eq_graphene_hamiltonian}), and $c^\dagger_{k\sigma}$ is the corresponding creation operator with  spin $\sigma$.
$|Ai\rangle$ represents the $\pi$ electron state localized at the site $i$($=2$ or $3$) of sublattice $A$.
$\delta(\varepsilon)$ is the Dirac delta function.
These operators are related to the original one $a_{i\sigma}$ by
\begin{eqnarray}
	\int\mathrm{d}\varepsilon a_{\varepsilon i\sigma} &=& a_{i\sigma}. 
\end{eqnarray}
The anti-commutation relations of these operators are calculated as $\left\{a_{\varepsilon i\sigma},a^\dagger_{\varepsilon' j\sigma}\right\}=\delta\left(\varepsilon-\varepsilon'\right)g_{ij}(\varepsilon)$, $\left\{a_{\varepsilon i\sigma},a_{\varepsilon' j\sigma}\right\}=0$,
and $\left\{a^\dagger_{\varepsilon i\sigma},a^\dagger_{\varepsilon' j\sigma}\right\} =0$, where $g_{ij}(\varepsilon)$ is introduced as
\begin{eqnarray}
	g_{ij}(\varepsilon)&=&\sum_k\langle k|Ai\rangle\langle Aj|k\rangle\delta\left(\varepsilon-\varepsilon_k\right).
\end{eqnarray}
The diagonal element $g_{ii}(\varepsilon)$ corresponds to the LDOS at site $i$ of sublattice $A$.
Owning to the twofold symmetry in the present system, the LDOS at the site $i=2$ and $3$ are identical: $g_{22}(\varepsilon)=g_{33}(\varepsilon)$.
The off-diagonal elements satisfy $g_{23}(\varepsilon)=g_{32}(\varepsilon)$, since the wave functions can be chosen to be real.
Then, from $a_{\varepsilon2\sigma}$ and $a_{\varepsilon3\sigma}$, the following operators which satisfy the usual anti-commutation relations are obtained;
\begin{eqnarray}
	c_{\varepsilon\sigma e} &=& \frac{1}{N^{1/2}_e(\varepsilon)}\left(a_{\varepsilon2\sigma}+a_{\varepsilon3\sigma}\right),\label{eq_operator_even}\\
	c_{\varepsilon\sigma o} &=& \frac{1}{N^{1/2}_o(\varepsilon)}\left(a_{\varepsilon2\sigma}-a_{\varepsilon3\sigma}\right),
\end{eqnarray}
where $e$ and $o$ denote even and odd, respectively, and
\begin{eqnarray}
	N_e(\varepsilon) &=& g_{22}(\varepsilon)+g_{23}(\varepsilon)+g_{32}(\varepsilon)+g_{33}(\varepsilon),\nonumber\\
	N_o(\varepsilon) &=& g_{22}(\varepsilon)-g_{23}(\varepsilon)-g_{32}(\varepsilon)+g_{33}(\varepsilon).\label{eq_normalization}
\end{eqnarray}
With these operators, the hybridization term (\ref{eq_hybridization}) is expressed as 
\begin{eqnarray}
	H_{\rm{hyb}} &=& \sum_\sigma\int\mathrm{d}\varepsilon N^{1/2}_e(\varepsilon)\left(Vc^\dagger_{\varepsilon\sigma e}d_\sigma+\mathrm{h.c.}\right).\label{eq_energy_rep_hyb}
\end{eqnarray}
Since $N_e(\varepsilon)$ is essentially the LDOS of $\pi$ electrons at the defect, this expression shows the importance of the LDOS at the defect, not the total DOS.
Note that only one channel of the conduction electron $c_{\varepsilon\sigma e}$ hybridizes with the $sp^2$ orbital.
This is because the $sp^2$ orbital hybridizes equally with the $\pi$ orbitals at sites $2$ and $3$ as in Fig.~\ref{fig_distortion}. 
When the hybridization is not equal, both even and odd states are coupled to the $sp^2$ orbital.

Using eqs.~(\ref{eq_graphene_hamiltonian}), (\ref{eq_operator_energy_site_rep}), and (\ref{eq_operator_even}), it can be shown that $c_{\varepsilon\sigma e}$ satisfies the commutation relation $\left[H_{\rm gra},c^\dagger_{\varepsilon\sigma e}\right]=(\varepsilon-\mu)c^\dagger_{\varepsilon\sigma e}$. 
Thus, neglecting the other channels of the conduction electrons, we assume that the conduction electron term (\ref{eq_graphene_hamiltonian}) can be written as [31-33]%~\cite{Jones1987, Sakai1992, Affleck1995}
\begin{eqnarray}
	H_{\rm{gra}}&=&\sum_{\sigma}\int\mathrm{d}\varepsilon\left(\varepsilon-\mu\right)c^\dagger_{\varepsilon\sigma e}c_{\varepsilon\sigma e}.\label{eq_energy_rep_pi}
\end{eqnarray}
Since decoupled channels of conduction electrons are neglected, this model can be used only when the defect contribution is concerned.
The transformation above has also been justified in ref.~\cite{Bulla1997} by integrating out the conduction electron degrees of freedom in the functional integral representation.

%%%%%%%%%%%%%%%%%%%%%%%%%%%%%%%%%%%%%%%%%%%
\section{Role of the Localized State of $\pi$ Electrons at the Defect: Numerical Renormalization Group Study }\label{sec_localized_state}
\begin{figure}
	\centering
		\includegraphics[width=6cm]{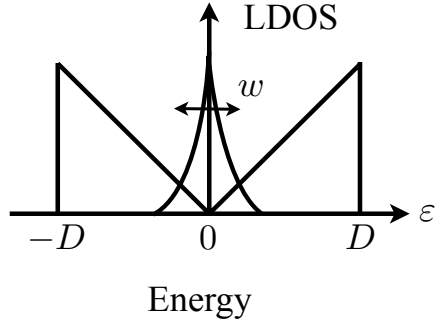}
	\caption{Local density of states of conduction electrons in the model in \S\ref{sec_localized_state} as a function of energy $\varepsilon$, where $D$ is a cut-off energy.
		A peak due to a localized state of $\pi$ electrons is included, where $w$ is a width of the peak.}
	\label{fig_kondo_ldos_loc}
\end{figure}

In this section, we discuss the effect of the localized state of $\pi$ electrons on the Kondo effect.
The localized state of $\pi$ electrons causes a peak in the LDOS on the defect as a function of energy at the charge neutrality point.
To take into account this peak of the LDOS, we assume a model shown in Fig.~\ref{fig_kondo_ldos_loc},
\begin{eqnarray}
	N_e(\varepsilon)=|\varepsilon|+\frac{\lambda}{2w}\mathrm{e}^{-|\varepsilon|/w},
\end{eqnarray}
where $\lambda$ and $w$ are the intensity and the width of the peak, respectively.
The second term becomes a Dirac delta function in the limit $w\to0$.
In the case of low density of defects, the LDOS at the defect is a Dirac delta function with zero width.
On the other hand, in the case of finite concentration of defects, the LDOS can have a finite width owning to overlaps of the wave functions of the localized states on different defects~\cite{Peres2006, Pereira2008}.
Assuming finite widths of the LDOS, we analyze this model by the NRG method.
The $\mu=0$ case is studied, where the effect of the peak of the LDOS is most significant.
\begin{figure}
	\centering
	\includegraphics[width=16cm]{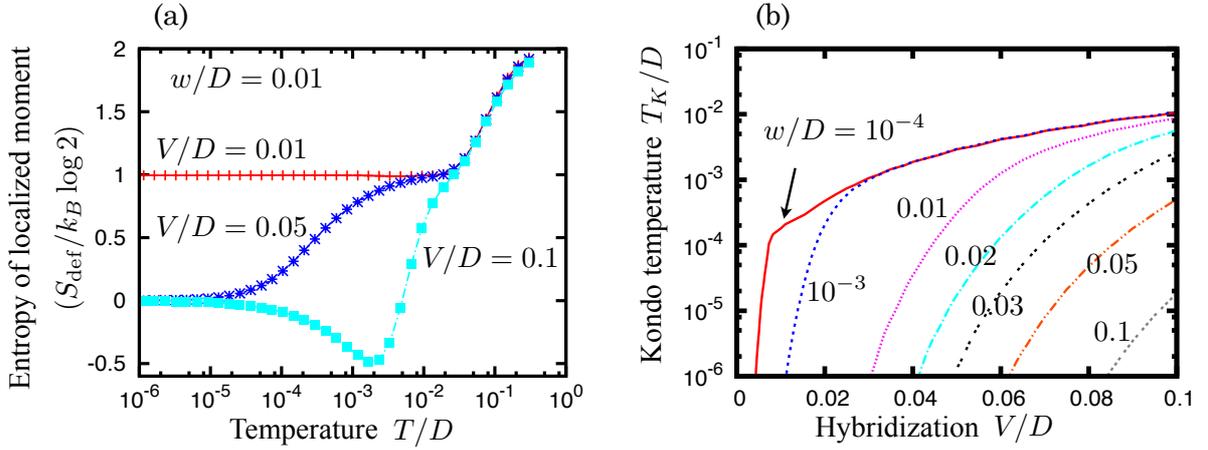}
	\caption{(a) Temperature dependence of entropy of the electron on the defect $sp^2$ orbital, $S_{\rm{def}}/\log2$, in the presence of a peak in the local density of states of $\pi$ electrons with the width $w/D=0.01$ at the chemical potential $\mu=0$, for the hybridization $V/D=0.01$, $0.05$, and $0.1$.
	(b) Kondo temperature $T_K/D$ as a function of hybridization $V/D$ for several widths $w/D$.
[Reprinted from Taro Kanao, Doctor Thesis, Department of physics, Graduate school of science, The university of Tokyo, Tokyo (2014). ]}
\label{fig_kondo_localized_state}
\end{figure}

Figure~\ref{fig_kondo_localized_state}(a) shows a temperature dependence of entropy of the electron on the defect $sp^2$ orbital, $S_{\rm{def}}/k_B\log2$, for $V/D=0.01$, $0.05$, and $0.1$ in the presence of finite peak width $w/D=0.01$.
In this NRG calculation, the parameter of the logarithmic intervals, $\Lambda$, is set to be $\Lambda=2$, and 300 states are retained in every step of diagonalization.

It is found that the low-energy state ($\sim10^{-6}D$) changes from the unscreened localized moment state ($S_{\rm{def}}/k_B\log2 =1$) to the Kondo-Yosida singlet state ($S_{\rm{def}}/k_B\log2 =0$) with increasing hybridization from $V/D=0.01$ to $0.05$.
Since $D$ is about 8 eV, $V$ is of the order of $0.1$ eV.
Namely, this Kondo effect can occur with a realistic smaller value of hybridization between the $sp^2$ orbital at the defect and the $\pi$ electrons.

In the case of $V/D=0.1$, $S_{\rm{def}}$ is negative in a certain range of temperature.
Note that $S_{\rm def}$ is defined as the deference between the entropy in the presence of defect and that without defect.
This is due to the large hybridization comparing with the band width.
The similar behavior has been known in the narrow-band Anderson model~\cite{Hofstetter1999}.

In order to understand the importance of localized $\pi$ orbital in detail, we also discuss the width of localized $\pi$ orbital dependence of the Kondo temperature. 
In Fig.~\ref{fig_kondo_localized_state}(b), the Kondo temperatures $T_K/D$ as a function of $V/D$ for several $w/D$ are shown.
It is found that the Kondo temperature is sensitive to the width: With smaller $w/D$, $T_K$ becomes larger owning to the larger LDOS at the Fermi energy.

%%%%%%%%%%%%%%%%%%%%%%%%%%%%%%%%%%%%%
\section{Effects of Magnetic Field}\label{sec_orbital_effect}
\begin{figure}
	\centering
	\includegraphics[width=16cm]{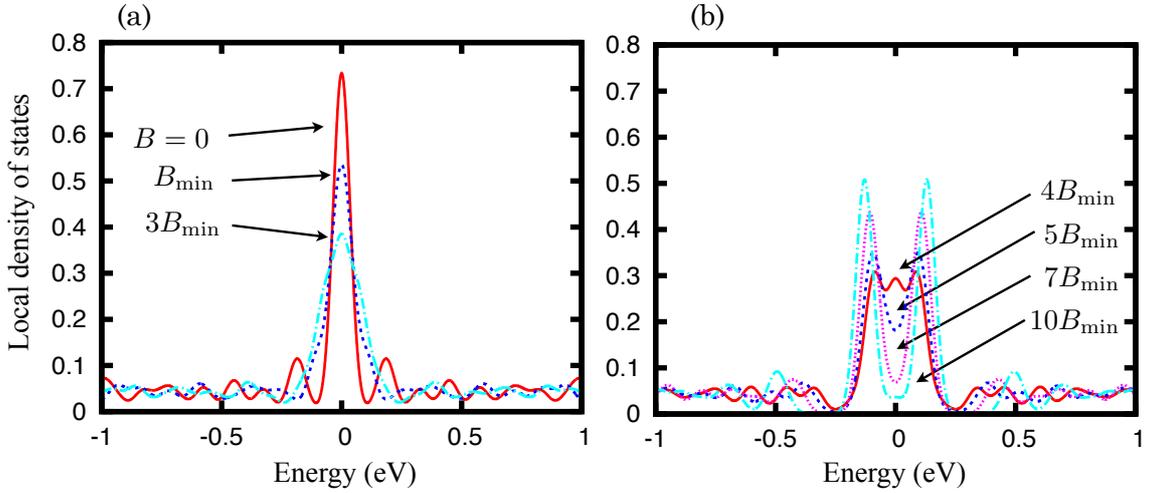}
	\caption{Local density of states of the $\pi$ electrons at a point defect as a function of energy in the magnetic filed of (a) $B/B_{\rm{min}}=0$, $1$, and $3$ and (b) $B/B_{\rm{min}}=4$, $5$, $7$, and $10$.
	$B_{\rm{min}}$ is determined by the system size.
	In the present calculation, $B_{\rm{min}}\simeq21.2$ T.
	[Reprinted from Taro Kanao, Doctor Thesis, Department of physics, Graduate school of science, The university of Tokyo, Tokyo (2014). ]}\label{fig_kondo_orbital_field}
\end{figure}
As mentioned in the introduction, the negative magnetoresistance has been observed experimentally in this Kondo effect~\cite{Chen2011}.
In the usual Kondo effect, the Zeeman effect on the localized magnetic moment causes a negative magnetoresistance, because it suppresses the formation of the Kondo-Yosida singlet.
The Kondo-Yosida singlet is suppressed when the energy scales of the magnetic field and the Kondo temperature become the same order, which leads to a characteristic magnetic field in the negative magnetoresistance.
In the Kondo effect in graphene with defects, however, the experimentally observed characteristic magnetic field is smaller than the energy scale of its Kondo temperature by one order of magnitude.
This means that this Kondo effect (Kondo temperature) is more sensitive to the magnetic field than the usual one.
However, the origin of this behavior has not been understood.

In the previous section, we have shown that the Kondo screening is assisted by the peak of the LDOS of $\pi$ electrons which is caused by the defect scattering.
Since the Kondo temperature sensitively depends on the height of the peak, small modulation of the localized $\pi$ orbital by the magnetic field will cause appreciable change in the Kondo screening.
Thus, to clarify the modulation due to the orbital motion of the $\pi$ electrons in the magnetic field, as a first step, we solve the nearest-neighbor tight-binding model for the $\pi$ states of graphene in the magnetic field with a single point defect.
In this case, the Hamiltonian is given as
\begin{eqnarray}
H_{\rm{gra}}^{\prime} = -t\sum_{\langle ij\rangle}{}^{'}\left(e^{i\theta_{ij}}a^\dagger_{i}b_{j}+\rm{h.c.}\right),
\end{eqnarray}
where ${}^{\prime}$ denotes the presence of the defect, and $a_{i}$ and $b_{j}$ are annihilation operators of a $\pi$ electron at site $i$ and site $j$ on graphene, and $\theta_{ij}$ is a Peierls phase with the \lq\lq string gauge"~\cite{Hatsugai1999}.
Under the periodic boundary condition, the possible magnetic fields are restricted to $B=nB_{\rm{min}}$ with $n=0,1,2,\cdots$, where $B_{\rm{min}}$ is determined by the system size.
In the present calculation, $61^2$ unit cells are used, which corresponds to $B_{\rm{min}}\simeq21.1$ T.

Figures~\ref{fig_kondo_orbital_field} show the LDOS at the defect as a function of energy in several magnetic fields.
We find that the height of the peak decreases and the width increases for relatively small $B$ as shown in Fig.~\ref{fig_kondo_orbital_field}(a).
At higher magnetic fields, the peak splits into two peaks as shown in Fig.~\ref{fig_kondo_orbital_field}(b).
These results show that the localized $\pi$ orbital becomes broad and the LDOS of the Dirac point (the Fermi level) decreases by the orbital magnetic field.
Since the Kondo temperature is strongly sensitive to the width of the localized $\pi$ orbital, it is expected that the Kondo temperature decreases drastically by the orbital magnetic field.

%%%%%%%%%%%%%%%%%%%%%%%%%%%%
\section{Summary}\label{sec_kondo_summary}
In conclusion, we discussed the role of localized $\pi$ orbital on the defect-induced Kondo effect in graphene on the basis of a numerical renormalization group study.  
We found that the localized $\pi$ orbital assists the Kondo effect due to the sp$^2$ orbital, and the Kondo temperature is sensitive to the broadening of the localized $\pi$ orbital.
Secondly, in order to clarify the mechanism of the magnetic sensitive Kondo effect, as a first step, we study the effect of the orbital magnetic field on the localized $\pi$ orbital by the tight-binding model with the Peierls phase.
We also found that the LDOS of the localized $\pi$ orbital at the fermi level decreases as the magnetic field increases.
It is suggested the the defect induced Kondo effect is strongly sensitive to the orbital magnetic field.

\subsection*{Acknowledgments}
This work has been supported by a Grant-in-Aid for Scientific Research A on "Dirac Electrons in Solids" (No.24244053), and for Scientific Research on Innovative Areas "Ultra Slow Muon Microscope" (No.23108004) from the Ministry of Education, Culture, Sports, Science and Technology, Japan.
One of authors (H.M.) is supported by a Grant-in-Aid for Scientific Research from the Japan Society for the Promotion of Science (No.25220803).

\end{document}